\documentclass[manuscript]{aastex}

\shorttitle{SEARCH FOR GAMMA RAY BURSTS}
\shortauthors{AIELLI ET AL.}

\begin{document}


\title{SEARCH FOR GAMMA RAY BURSTS WITH THE ARGO-YBJ DETECTOR IN SCALER MODE}


\author{G. Aielli\altaffilmark{1,2}, 
C. Bacci\altaffilmark{3,4}, 
F. Barone\altaffilmark{5,6},
B. Bartoli\altaffilmark{6,7}, 
P. Bernardini\altaffilmark{8,9}, 
X.J. Bi\altaffilmark{10},
C. Bleve\altaffilmark{8,9},
P. Branchini\altaffilmark{4}, 
A. Budano\altaffilmark{4},
S. Bussino\altaffilmark{3,4}
% ...
}

% ...
\altaffiltext{1}{Dipartimento di Fisica dell'Universit\`a ``Tor Vergata''
                 di Roma, via della
                 Ricerca Scientifica 1, 00133 Roma, Italy.}
\altaffiltext{2}{Istituto Nazionale di Fisica Nucleare, Sezione di Tor 
        Vergata, via della Ricerca Scientifica 1, 00133 Roma, Italy.}
\altaffiltext{3}{Dipartimento di Fisica dell'Universit\`a ``Roma Tre'' di 
                 Roma, via della Vasca Navale 84,
                 00146 Roma, Italy.}
\altaffiltext{4}{Istituto Nazionale di Fisica Nucleare, Sezione di
                   Roma3, via della Vasca Navale 84, 00146 Roma, Italy.}
\altaffiltext{5}{Dipartimento di Scienze Farmaceutiche dell'Universit\`a
                 di Salerno, via Ponte Don Melillo, 
                 84084 Fisciano (SA), Italy.}
\altaffiltext{6}{Istituto Nazionale di Fisica Nucleare, Sezione di
                 Napoli, Complesso Universitario di Monte
                 Sant'Angelo, via Cinthia, 80126 Napoli, Italy.}
\altaffiltext{7}{Dipartimento di Fisica dell'Universit\`a di Napoli, 
                 Complesso Universitario di Monte
                 Sant'Angelo, via Cinthia, 80126 Napoli, Italy.}
\altaffiltext{8}{Dipartimento di Fisica dell'Universit\`a del Salento, 
                 via per Arnesano, 73100 Lecce, Italy.}
\altaffiltext{9}{Istituto Nazionale di Fisica Nucleare, Sezione di
        Lecce, via per Arnesano, 73100 Lecce, Italy.}
\altaffiltext{10}{Key Laboratory of Particle Astrophyics, Institute of High
                  Energy Physics, Chinese Academy of Science,
                  P.O. Box 918, 100049 Beijing, P.R. China.}
% ...

\begin{abstract}

We report on the search for Gamma Ray Bursts (GRBs) in the energy range 
1-100 GeV in coincidence with the prompt emission detected by satellites 
using the Astrophysical Radiation with Ground-based Observatory at YangBaJing
(ARGO-YBJ) air shower detector.
Thanks to its mountain location (Yangbajing, Tibet, P.R. China, 4300 m a.s.l.),
active surface ($\sim$6700 m$^2$ of Resistive Plate Chambers), and
large field of view ($\sim$2 sr, limited only by the atmospheric
absorption), the ARGO-YBJ air shower detector is particularly suitable for the 
detection of unpredictable and short duration events such as GRBs. 
The search is carried out using the ``single particle technique'',
i.e. counting all the particles hitting the detector without measurement of
the energy and arrival direction of the primary gamma rays.

Between 2004 December 17 and 2009 April 7, 81 GRBs detected by 
satellites occurred within the field of view of ARGO-YBJ 
(zenith angle $\theta \leq 45^\circ$). It was possible to examine
62 of these for $>$ 1 GeV counterpart in the 
ARGO-YBJ data finding no statistically significant emission.
With a lack of detected spectra in this energy range  
fluence upper limits are profitable, especially when the redshift is known and 
the correction for the extragalactic absorption can be considered.
The obtained fluence upper limits reach values as low as $10^{-5}$ erg cm$^{-2}$
in the 1-100 GeV energy region.
\\
Besides this individual search for a higher energy counterpart, a 
statistical study of the stack of all the GRBs both in time 
and in phase was made, looking for a common feature in the GRB high 
energy emission. No significant signal has been detected.

\end{abstract}





\section{INTRODUCTION}

Over the past ten years, a considerable effort has been made to study 
Gamma Ray Bursts (GRBs).
The solution of the puzzle about their distance 
by the BeppoSAX satellite in 1997 with the detection of the first 
afterglow \citep{Cos97} was followed by a small fleet of satellites 
(Swift, HETE, AGILE, and now the Fermi Gamma Ray Space Telescope) that 
performed an extensive study in the keV-MeV energy range.
At ground level, air shower detectors can contribute to the study of the 
energy region $\geq$ 1 GeV using the 
``single particle technique'' \citep{Ver00}, 
i.e. counting all the particles hitting the detector without measurement 
of the energy and arrival direction of the primary gamma rays. 
\\
The processes leading to the detection of  Very High Energy (VHE) radiation
(extending between 1 GeV and 1 TeV) from a GRB start with the acceleration 
of the parent particles. 
Since the models predicting the production of gamma rays in this energy 
range are both leptonic and hadronic, the maximum energy of electrons and 
protons is linked to the fireball parameters \citep{Fox06}.
Once the parent particles have been accelerated, the VHE gamma rays can
result from several production  processes - mainly inverse Compton
scattering by electrons, electron and proton 
synchrotron emission, and photon-pion production - each of them included in
a wide variety of theoretical models with different hypotheses on the
production region, giving different features of the emitted signal
(see for example the review article by \citet{Mes06} and references therein).
Once produced, the VHE radiation suffers absorption either in the source 
itself or in the Extragalactic
Background Light (EBL) before reaching the Earth \citep{Sal98,Tot00}.
The detected signal is thus the overlap of the previous mechanisms, that 
are difficult  to disentangle; nevertheless the measurements in this energy
range can be used as a test for the competing models.
In particular, the detection of VHE gamma rays and the measurement of 
the cutoff energy in the GRB spectrum could provide valuable clues to 
the baryon content, Lorentz factor, and ambient magnetic field of the 
relativistic fireball.
No cutoff energy has been detected so far by satellites up to 1 GeV, 
forcing the search into more energetic regions.
The energy range between 1 and 100 GeV is particularly interesting since 
the absorption of high energy gamma rays by the EBL becomes relevant 
above this region, limiting the possibility of detection to nearby objects  
while most of the observed GRBs occurs at large redshifts.
Anyway the knowledge of the EBL is still poor 
and the problem of the opacity of the Universe to gamma rays in the
sub-TeV energy range still open (e.g. \citealp{Alb08,Ste08}).
\\
In the 1-100 GeV energy region, the detection by EGRET of only 
3 GRBs \citep{Cat97} during 7 years of observations 
indicates that their spectra are usually soft.
Recently, the Large Area Telescope (LAT) on board the Fermi 
Gamma Ray Space Telescope
announced the detection of more than 10 photons above 1 GeV from 
GRB080916C \citep{Taj08} and of emission up to 3 GeV from 
GRB081024B \citep{Omo08}
(unfortunately both these events were below the horizon of our detector).
At higher energies, hints ($\sim$ 3$\sigma$) of emissions detected at 
ground level have been reported by Milagrito for GRB970417A (E $>$ 650
GeV) \citep{Atk00} and by the GRAND array for GRB971110 
(E $>$ 10 GeV) \citep{Poi03}.
A marginal emission for GRB920925C was also reported by HEGRA AIROBICC 
(E $>$ 20 TeV) \citep{Pad98}, while the Tibet Air Shower array found an 
indication of 10 TeV emission in a stacked analysis of 57 bursts \citep{Ame96}.
\\
Forty years after their discovery, and more than ten years after the detection 
of the first afterglow by BeppoSAX, the physical origin of 
the enigmatic GRBs is still under debate. 
The scarcity of information generates a confused
situation, allowing a great variety of very different models.
In these conditions, and mainly in the $>$ 1 GeV energy region, any result 
could be of great importance to approach the solution of the GRB mystery.
\\
In this paper, the search for emission in the 1-100 GeV range 
in coincidence with the prompt emission detected by satellites 
is presented for several GRBs.

\section{THE DETECTOR}


The Astrophysical Radiation with Ground-based Observatory at YangBaJing
(ARGO-YBJ) is an extensive air shower detector located at an altitude of
4300 m a.s.l. (corresponding to a vertical atmospheric depth of 
606 g cm$^{-2}$) at the Yangbajing Cosmic Ray Laboratory (30.11$^{\circ}$N, 
90.53$^{\circ}$E).
The detector is composed of a single layer of Resistive Plate Chambers (RPCs), 
operated in streamer mode \citep{Aie06} and grouped
into 153 units called ``clusters'' (5.7$\times$7.6 m$^2$).
Each cluster is made up of 12 RPCs (1.225$\times$2.850 m$^2$) and
each RPC is read out by using 10 pads,
with dimensions 55.6$\times$61.8 cm$^2$, representing the space-time 
pixels of the detector.
The clusters are disposed in a central full coverage carpet (130 clusters,
$\sim$5600 m$^2$, $\sim$93\% of active surface) and a sampling guard
ring ($\sim$40\% of coverage) in order to increase the effective area and 
improve the core location reconstruction.
\\
The detector is connected to two independent data acquisition systems,
corresponding to the shower and scaler operation modes.
\\
In shower mode the arrival time and location of each
particle are recorded using the ``pads''.
The present trigger threshold is set to 20 fired pads,
corresponding to an energy threshold for photons of a few hundred GeV
and a trigger rate of $\sim$ 3.8 kHz.
\\
In scaler mode the total counts are measured every 0.5 s: for each cluster
the signal coming from its 120 pads,
representing the counting rate on a surface of $\sim$43 m$^2$, 
is added up and put in coincidence in a narrow time interval (150 ns),
giving the counting rates of $\geq$ 1, $\geq$ 2, $\geq$ 3, and $\geq$ 4 
pads, that are read by four independent scaler channels.
These counting rates are referred in the following respectively as 
$C_{\ge 1}$, $C_{\ge 2}$, $C_{\ge 3}$, and 
$C_{\ge 4}$, and 
the corresponding rates are $\sim$ 40 kHz, $\sim$ 2 kHz, $\sim$ 300
Hz, and $\sim$ 120 Hz.
A detailed description of the detector performance can be found in
\citet{Pds08} and references therein.
\\
The installation of the whole detector was completed in spring 2007,
but since the clusters work independently, physical studies started
as the installation began, with the active area
increasing with time. 
Although the single particle technique does not provide 
information about the energy and arrival direction 
of the primary gamma rays, it allows the energy threshold to be lowered to
$\sim$ 1 GeV, thus overlapping the highest energies investigated by satellite 
experiments. 
Moreover, in our application of this technique to the ARGO-YBJ experiment 
\citep{Aie08}, with four measurement channels sensitive to different energies,
in case of positive detection valuable information on the high energy
spectrum slope and possible cutoff may be obtained.
\\
Since for the GRB search in scaler mode the authentication is only given
by the satellite detection, the stability of the detector and the
probability that it mimics a true signal are crucial and have to be carefully
investigated.
Details of this study are in \citet{Aie08}, 
together with the determination of the
effective area, upper limit calculation, and expected sensitivity.
The results obtained show that the main influences on the counting rates are
given by the atmospheric pressure (barometric coefficient for modulation: 
-(0.9 - 1.2) \% mbar$^{-1}$, connected to the shower development in the atmosphere) 
and the detector temperature (thermal coefficient: 0.2 - 0.4\% $^{\circ}$C$^{-1}$, 
linked to the detector efficiency).
Due to the larger time scale variations for these two parameters with
respect to the GRB prompt phase duration, the search for GRBs can be carried 
out without any correction for environmental or instrumental effects, 
since the relevant distributions for single clusters are Poissonian.
More significant is the correlation between the clusters due to the 
probability that some of the counts in different clusters are given by the same
events: the effect in this case is to widen background fluctuations,
reducing the sensitivity.
\\
The GRB search can be performed in both shower and scaler modes; in this paper 
only the results obtained with the latter are presented and discussed.




\section{GRB SEARCH}

\begin{figure}
\plotone{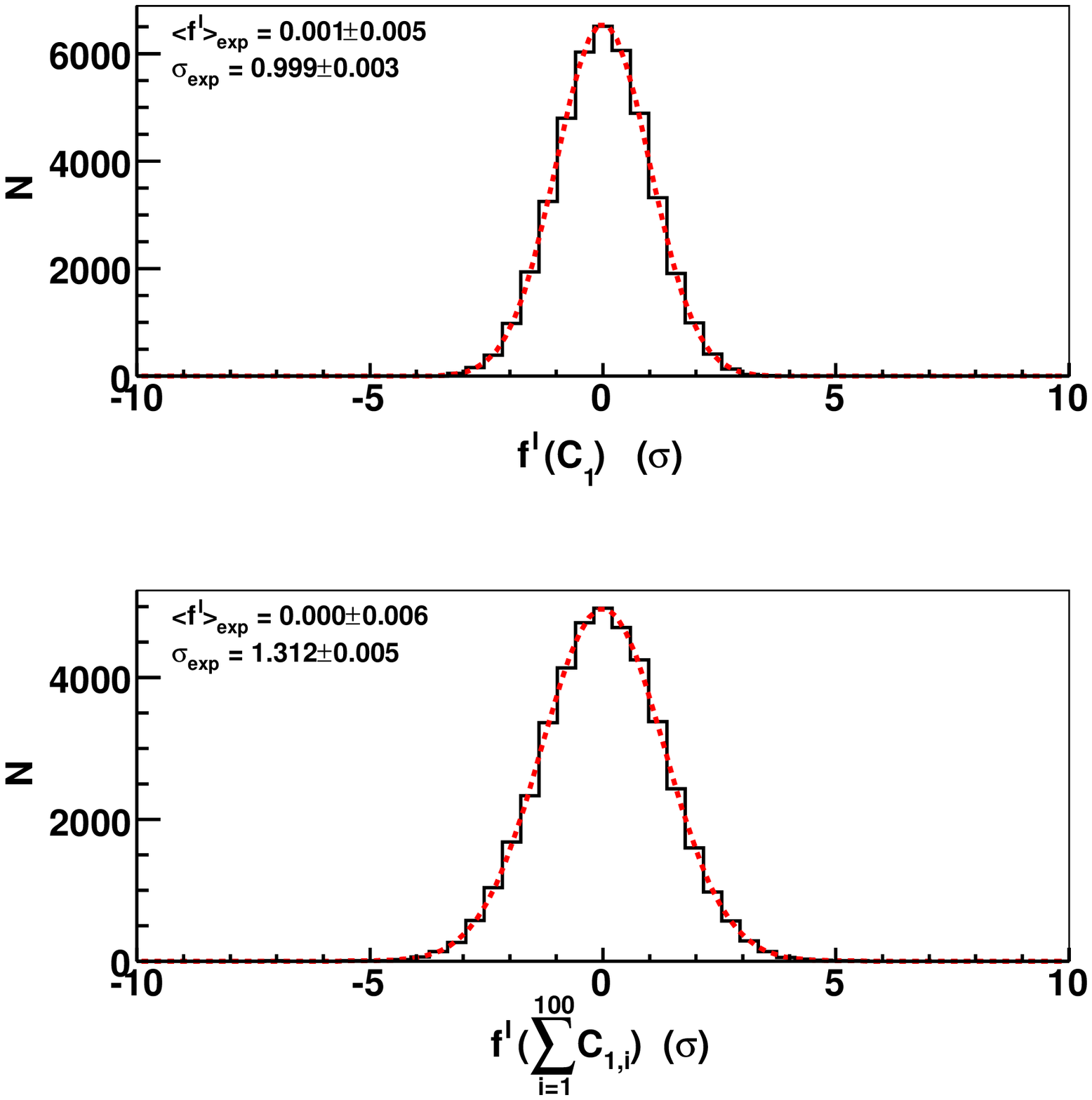} 
\caption{Experimental distribution of the normalized excesses 
of signal over background for GRB060121.
Top: $C_1$ channel for a typical cluster compared with a 
Gaussian fit; bottom: sum of the 100 active clusters.}
\label{fig:fdistrib}
\end{figure}

\begin{figure}
\plotone{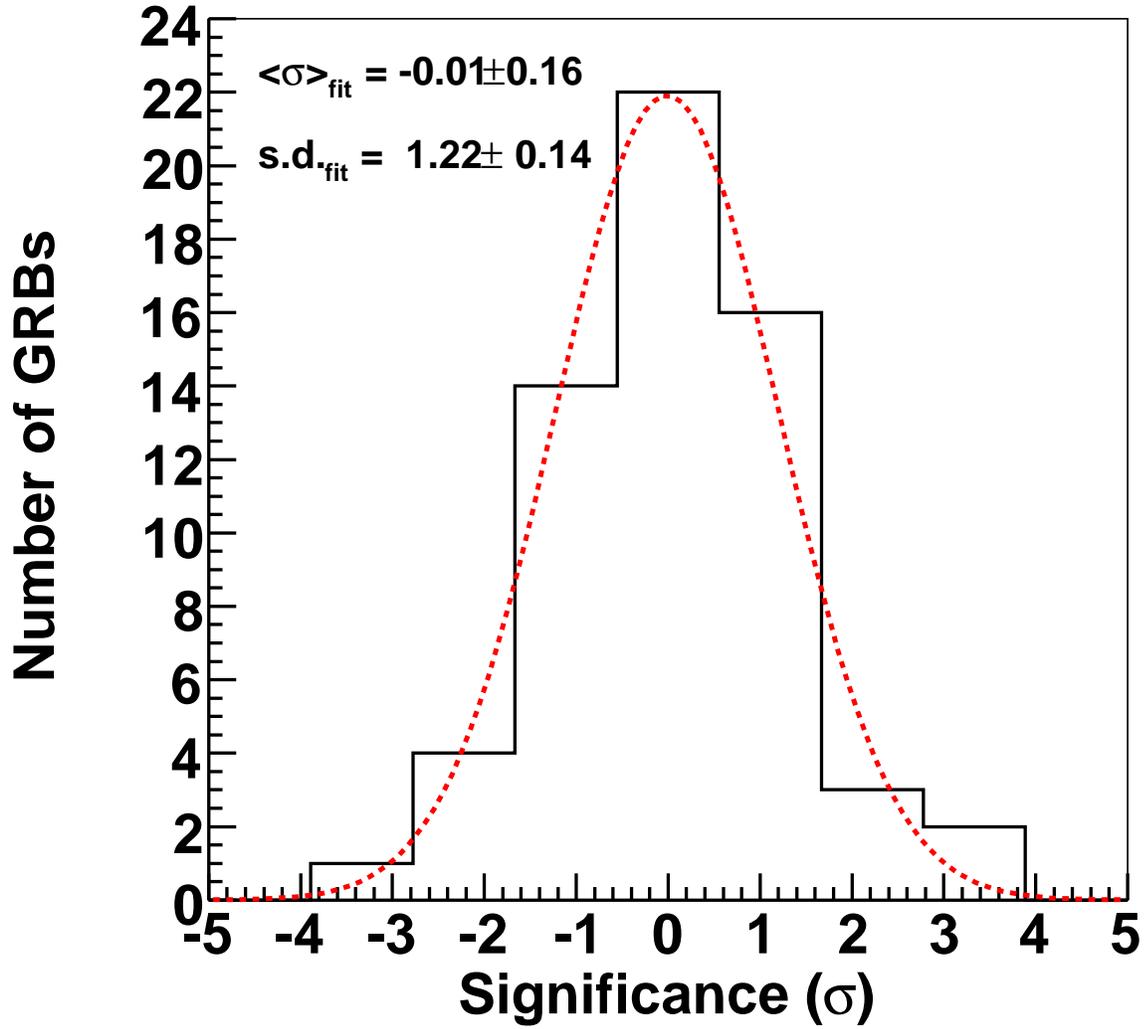}
\caption{Distribution of the statistical significances of
  the 62 GRBs with respect to background fluctuations,
  compared with a Gaussian fit.}
\label{fig:sigma}
\end{figure}

\begin{figure}
\plotone{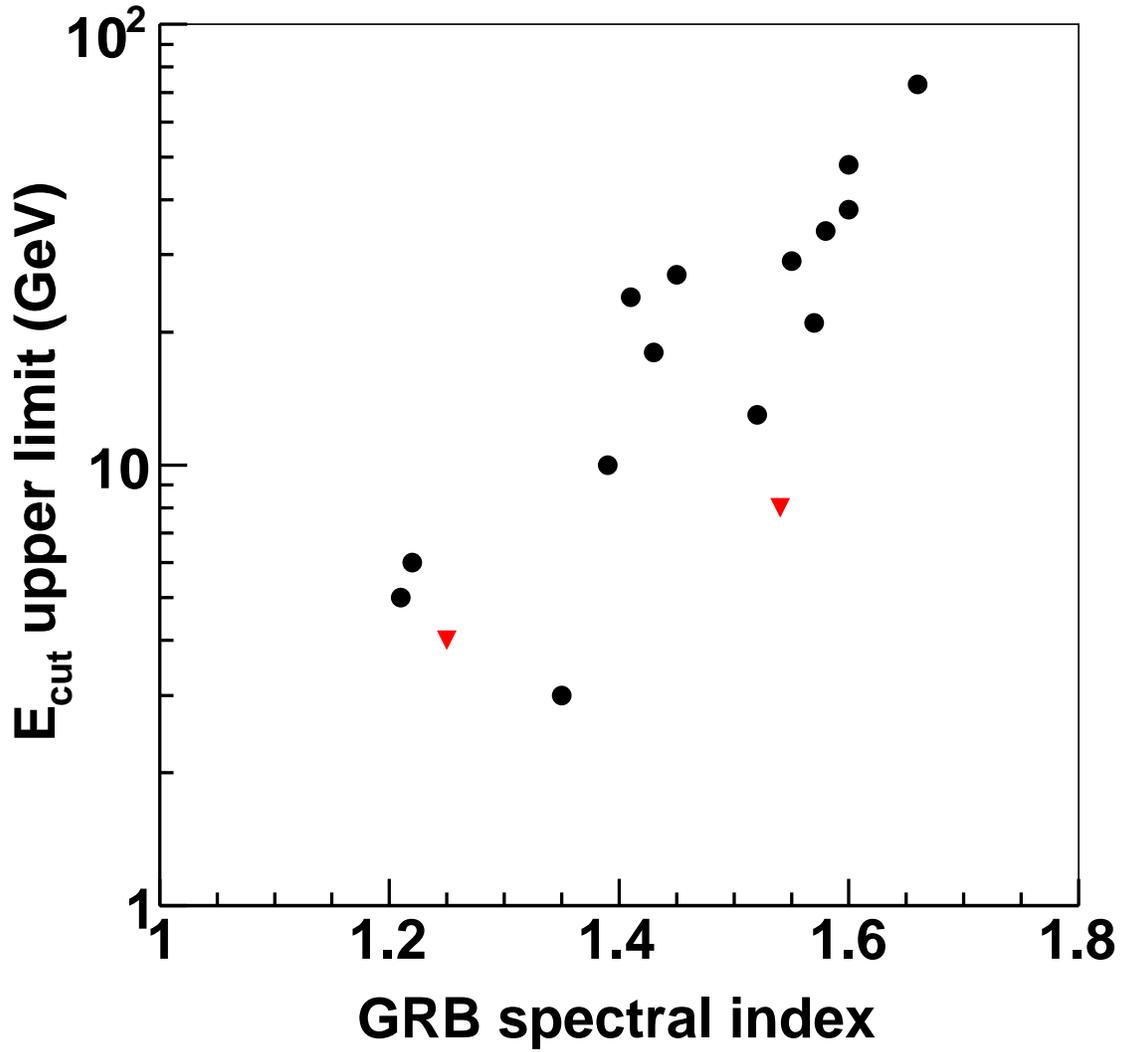}
\caption{Cutoff energy upper limits as a function of the spectral
index obtained by extrapolating the measured keV spectra.
The values represented by the triangles are obtained taking into account 
extragalactic absorption.} 
\label{fig:ecut}
\end{figure}

\begin{figure}
\plotone{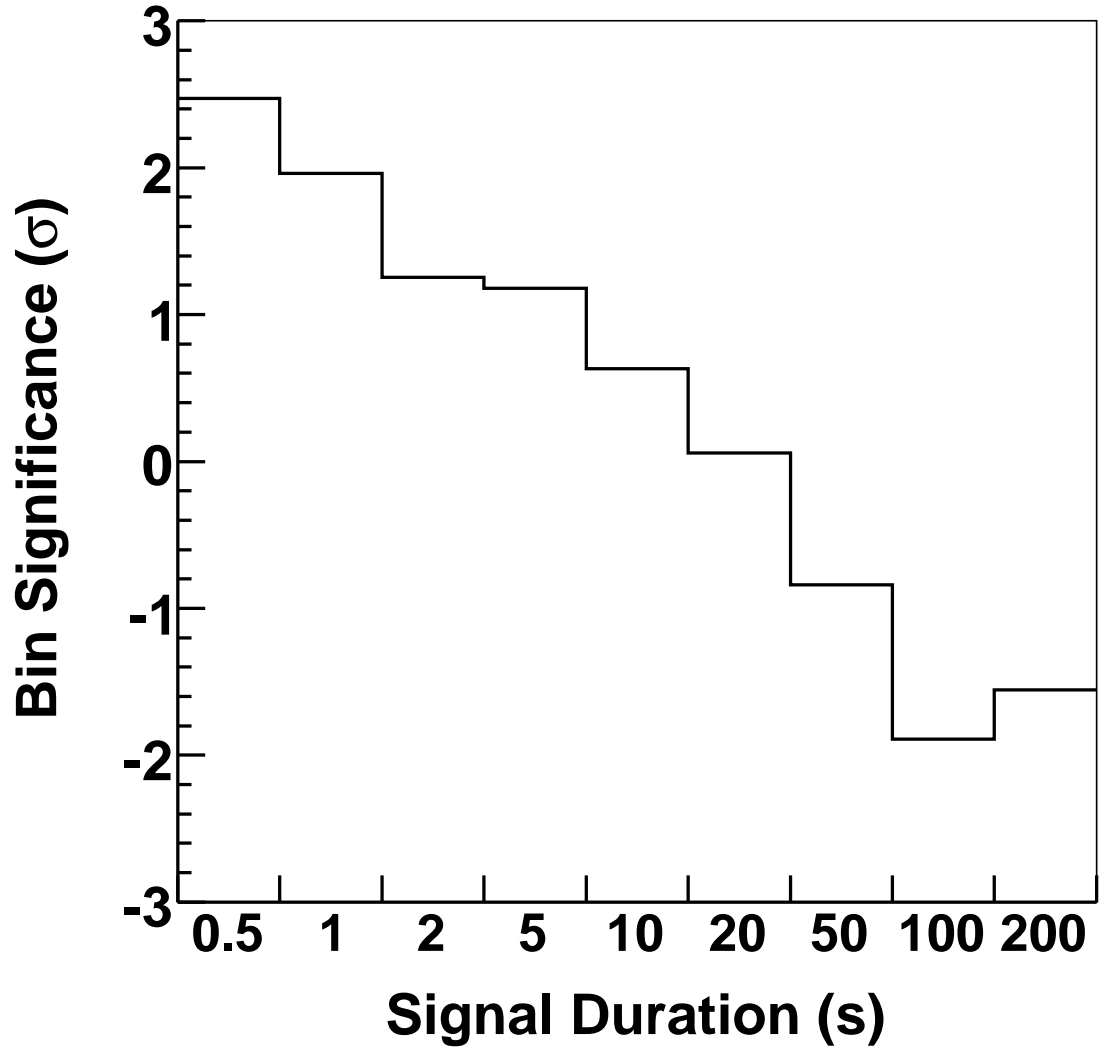}
\caption{Significances of GRBs stacked in time for durations 
between 0.5 and 200 s after the low energy trigger time T$_0$.}
\label{fig:stackt} 
\end{figure}

\begin{figure}
\plotone{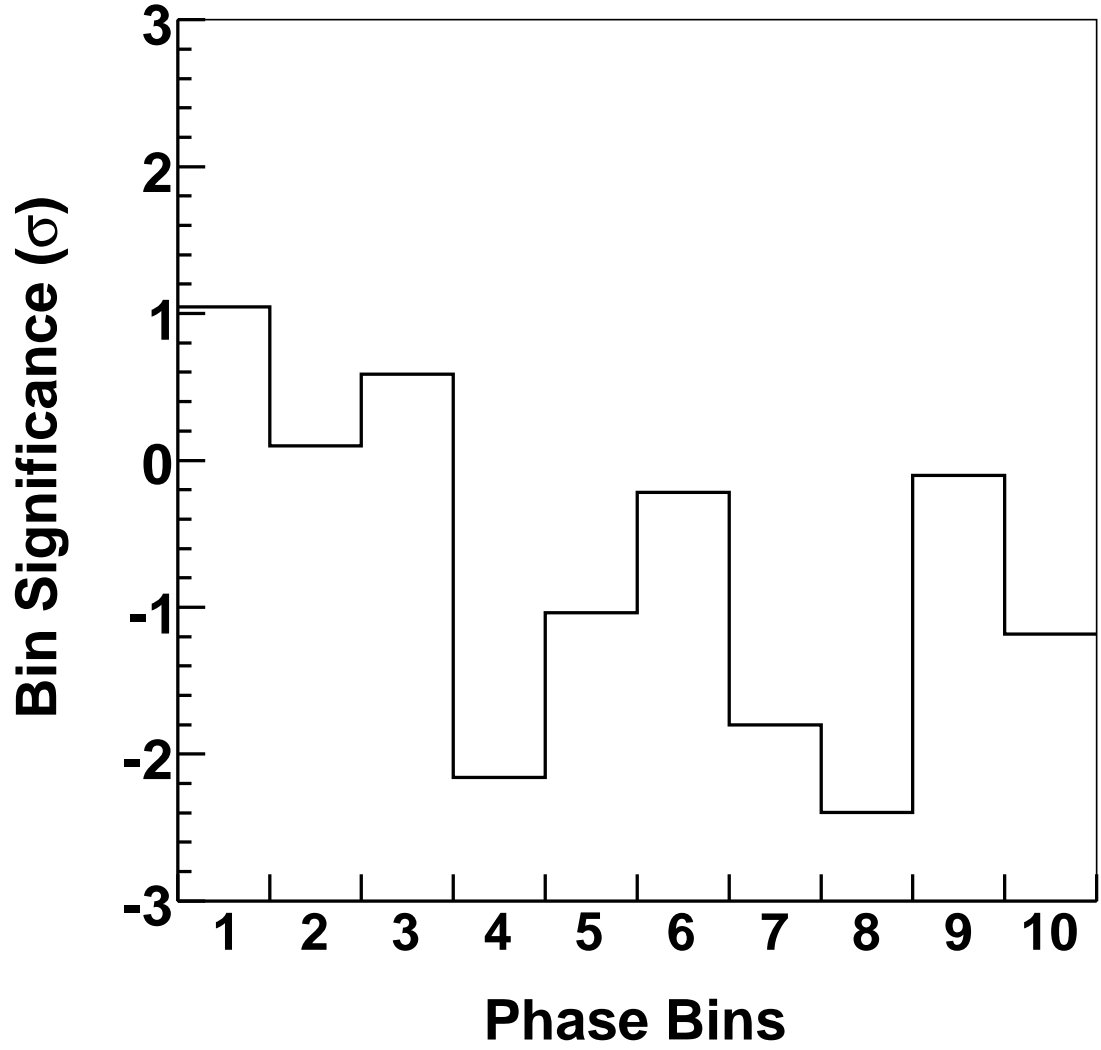}
\caption{Significances of GRBs with duration $\Delta t_{90} \ge 
5s$ stacked in phase (see text).}
\label{fig:stackp}
\end{figure}

Data have been collected from 2004 November   
(corresponding to the Swift satellite launch) to 2009 April, with
a detector active area increasing from $\sim$700 to $\sim$6700 m$^2$.
During this period, a total of 81 GRBs 
selected from the GCN Circulars
Archive\footnote{http://gcn.gsfc.nasa.gov/gcn3\textunderscore archive.html}
was inside the ARGO-YBJ field of view
(i.e. with zenith angle $\theta \leq 45^{\circ}$, limited only by the
atmospheric absorption); for 19 of these the 
detector and/or data acquisition were not active or not working properly.
The remaining 62 events were investigated by searching for a significant 
excess in the ARGO-YBJ data coincident with the satellite detection.
In order to extract the maximum information from the data, two GRB analyses
have been implemented:
\begin{itemize}
\item search for a signal from every single GRB;
\item search for a signal from the stack of all GRBs.
\end{itemize}
For both analyses, the first step is the data cleaning and check.
For each event, the 
Poissonian behaviour of the counting rates for multiplicities 
$\ge$1, $\ge$2, $\ge$3, $\ge$4 for all the clusters is checked using the
normalized fluctuation function:

\begin{equation}
f = (s-b)/\sigma, \; \; \; \; \; 
\sigma = \sqrt{b+b/20}
\label{ffunction}
\end{equation}
for a period of $\pm$12 h around the GRB trigger time.
In this formula, representing the significance of an excess compared to
background fluctuations, $s$ is the number of counts in a time interval
of 10 s, $b$ the average number of counts in 10 s over a time period of 
100 s before and after the signal, and $\sigma$ the
standard deviation, with about 400 independent samples per distribution.
The interval of 10 s has been chosen to avoid 
any systematic effect caused by environment and instrument
(such as atmospheric pressure and detector temperature variations).
The expected distribution of $f$ is the standard normal function;
all the clusters giving a distribution with measured $\sigma > 1.2$ or
with anomalous excesses in the tail $\sigma >$ 3 (i.e. $>$ 2\%) in at
least one multiplicity channel are discarded. 
This guarantees that our data fulfill the requirements on stability and
reliability of the detector.
The counting rates of the clusters surviving our quality cuts ($\sim$87\%) 
are then added up and the normalized fluctuation function

\begin{equation}
f' = (s'-b')/\sigma', \; \; \; \; \; 
\sigma' = \sqrt{b'+b'\frac{\Delta t_{90}[s]}{600}}
\label{ffunction2}
\end{equation}
is used to give the significance of the coincident on-source counts.
In this case $s'$ is the total number of counts in the $\Delta t_{90}$ time
interval given by the satellite detector (corresponding to the collection of
90\% of the photons) and $b'$ is the number of counts in a fixed time 
interval of
300 s before and after the signal, normalized to the $\Delta t_{90}$ time.
Due to the correlation between the counting rates of different clusters 
(given by the air shower 
lateral distribution), the distributions of the sum of the counts are
larger than Poissonian and this must be taken into account to calculate the
significance of a possible signal.
The statistical significance of the on-source
counts over the background is obtained again in an interval of 
$\pm$12 h around the GRB trigger time, using equation (17) of 
\citet{Li83}; a detailed analysis of the
correlation effect and detector stability on counting rates can be
found in \citet{Aie08}. 
The analysis can be carried out for the counting rates for all the 
multiplicities $\ge$1, $\ge$2, $\ge$3, $\ge$4, and 1, 2, 3, where the counting 
rates $C_i$ are obtained from the measured counting rates $C_{\ge i}$ using 
the relation:

\begin{equation}
C_i = C_{\ge i} - C_{\ge i+1} \; \; \; (i=1,2,3)
\end{equation}
As an example, figure \ref{fig:fdistrib} shows the $f'(C_1)$ distribution for a
single cluster and for the sum of all clusters for GRB060121; even if all
the single clusters show a Poissonian behaviour, with width $\sigma$ $\sim$
1, the correlation
effect on the sum of the counting rates of all clusters broadens the 
$f'(\sum C_{1,i})$ distribution ($\sigma$ $>$ 1). 
In the following, all the results are obtained using the counting
rate $C_1$, since it corresponds to the minimum primary energy in the
ARGO-YBJ scaler mode.
Figure \ref{fig:sigma} shows the distribution of the significances for the
whole set of 62 GRBs.
The Gaussian fit to this distribution gives a mean of (-0.01 $\pm$ 0.16)$\sigma$ and
standard deviation (1.22 $\pm$ 0.14).
No significant excess is shown; the maximum significance is obtained for
GRB080727C (3.52 $\sigma$), with a chance probability of 1.4\% taking into 
account the total number of GRBs analyzed.
\\
The fluence upper limits are then obtained in the 1-100 GeV energy range 
adopting a power law spectrum and considering the maximum number of 
counts at 99\% confidence level (c.l.), following equation (6) of
\citet{Hel82}.
For this calculation, two different assumptions are used for the power
law spectrum: a) extrapolation from the keV-MeV energy region using the 
spectral index measured by the satellite experiments; b) a
differential spectral index $\alpha$ = -2.5. 
Since the mean value of spectral indexes measured by EGRET
in the GeV energy region is $\alpha$ = -2.0 \citep{Din97}, we expect 
the true upper limits to lie between these two values.
For GRBs with known redshift, an exponential cutoff in the spectrum 
is considered in order to take into account the effects of extragalactic
absorption. 
The extinction coefficient is calculated using the values
given in \citet{Kne04}.
When the redshift is unknown, a value $z$ = 1 is adopted.
\\
Since the cutoff energy of GRBs is unknown, the following procedure is 
developed in order to determine an upper limit to this
energy at least for some GRBs. 
When using as the GRB spectrum the extrapolation of the spectral index
measured in the keV-MeV region by satellite experiments,
the extrapolated fluence is plotted together with our fluence upper limit
as a function of the cutoff energy $E_{cut}$. 
If the two curves cross in the 2-100 GeV energy range, the intersection 
gives the upper limit to the cutoff energy. 
For these GRBs we can state that their spectra do not extend over the obtained
$E_{cut}$ upper limit, with a 99\% c.l..
Figure \ref{fig:ecut} shows the cutoff energy upper limits as a function of
the spectral index for the 16 GRBs for which intersection occurs in
the quoted energy range.
For two of them (GRB050802 and GRB081028A) the knowledge of the redshift 
allows the estimation of extragalactic absorption.
\\
Table \ref{tab:ul_rs} lists all the information related to the 9 GRBs with 
known redshift; Table \ref{tab:ul} lists the same information for the 
remaining 53 GRBs. In both tables column (1) is the GRB name corresponding
to the detection date in UT (YYMMDD). Column (2) gives the satellite(s) that
detected the burst. Column (3) gives the burst duration $\Delta t_{90}$ 
as measured by the respective satellite. Column (4) gives the zenith angle 
in degrees with respect to the detector location. Column (5) reports the
spectral index: ``Band'' and ``CPL'' mean that the spectrum measured by the 
satellite is better fitted with a double power law \citep{Ban93} or a 
Cutoff Power Law, respectively. In this latter case no extrapolation of the 
spectrum to GeV energies has been considered. Column (6) gives the detector
active area for that burst. Our results are reported from column (7) to (10).
Column (7) gives the statistical significance of the on-source counts over
the background; columns (8) and (9) the 99\% confidence upper limits
on the fluence between 1 and 100 GeV for spectral cases a) and b),
respectively; column (10) the cutoff energy upper limit, if any. The 
additional column (11) in Table \ref{tab:ul_rs} gives the GRB redshift.
\\
The fluence upper limits obtained in the 1-100 GeV energy range depend
on the zenith angle, time duration and spectral index, reaching values
down to $10^{-5}$ erg cm$^{-2}$.
It is worth noticing that these values greatly depend on the energy
range of the calculation. If we consider our sensitivity in terms of the
expected number of positive detections, an estimate based on data from the
satellite CGRO and the Swift field of view gives a rate between 
0.1 and 0.5 per year \citep{Aie08},
which is comparable to similar evaluations for other experiments working in 
different energy regions (e.g. \citealp{Alb07}).
\\
A different analysis is performed supposing a common timing feature in all 
the GRBs. First, all the events during a time interval $\Delta$t 
(with $\Delta$t=0.5, 1, 2, 5, 10, 20, 50, 100, 200 s) after
T$_0$ (the low energy trigger time given by the satellite) for all the GRBs
are added up.
This is done in order to search for a possible cumulative high energy
emission with a fixed duration after T$_0$.
The resulting significances for the 9 time bins (figure \ref{fig:stackt}) 
show that there is no evidence of emission for any one of the durations
$\Delta$t.
Since the bins are not independent, the distribution  
of the significances of the 9 time bins is compared with random
distributions obtained for starting times different from T$_0$ in a time
interval of $\pm$12 h around the true GRB trigger time.
The resulting overall significance of the GRBs stacked in time with respect
to random fluctuations is -0.6 $\sigma$.
A second search is carried out to test the hypothesis that the high energy
emission occurs at a specific phase of the low energy burst, 
independently of the GRB duration.
For this study, all 53 GRBs with $\Delta t_{90} \ge 5\;s$ (i.e. belonging
to the ``long GRB'' population) have been added up in phase scaling 
their duration.
This choice has been made for both physical and technical reasons, adding
up the counts for GRBs of the same class and long enough to allow a
phase plot with 10 bins given our time resolution of 0.5 s.
Figure \ref{fig:stackp} shows the resulting significances for the 10 phase
bins; there is no evidence of emission at a certain phase, and the
overall significance of the GRBs stacked in phase (obtained adding up all
the bins) with respect to background fluctuations is -1.4 $\sigma$.
The search for cumulative effects by stacking all the GRBs either in fixed 
time durations or in phases of $\Delta t_{90}$ could enhance a possible 
signal, making it significant, even if the emission of each GRB is below the
sensitivity of the ARGO-YBJ detector.
In this case, less information could be given with respect to the single
GRB coincident detection, but we must consider that with the stacked
analysis we increase our sensitivity by increasing the number of
GRBs, while for the single GRB search we decrease our
sensitivity because of the increasing number of trials.

%






\section{DISCUSSION AND CONCLUSIONS}

The satellite-borne detectors have detected GRBs mostly in the sub-MeV
energy region.
However, several GRBs with emission beyond 100 MeV have been detected by
EGRET \citep{Sch92,Hur94,Cat97} and, recently, by AGILE \citep{Giu08} 
and by the LAT instrument on the Fermi Gamma Ray Space Telescope 
\citep{Taj08}.
These detections indicate that at least a fraction of GRBs, in addition
to sub-MeV photons, may also emit much higher energy photons, possibly
extending to the GeV-TeV region.
Within the standard fireball shock scenario, high energy photons can be 
produced in both internal \citep{Pac94,Mes94} and external \citep{Ree92} 
shocks, either by
the electron component through the inverse Compton process or by the 
proton component through synchrotron or photo-pion processes.
Due to the interaction with cosmic infrared background photons, most of the
high energy GRB photons are converted into electron-positron pairs,
thus limiting the distance over which they can travel.
A correlated detection of GRB high energy photons, associated to the
redshift measurement based on spectroscopic observations, could help 
determine the extension of the gamma ray horizon and shed light on the
problem of the universe ``transparency'' recently raised by the
observations of the HESS \citep{Aha06} and MAGIC \citep{Alb08} telescopes.
Proposed explanations of these observations refer to models for the EBL
evolution \citep{Ste06,Ste08} as well as the particle physics process of 
photon-ALP (Axion-Like Particle) oscillation \citep{Sik83,DeA08}.
Thus, the observation of GRB high energy photons is expected to provide
important information both on the physical conditions of the emission
region and on the interaction processes undergone by the photons while
traveling from the source.
\\
In this paper we have reported a study concerning the search for GeV photons
from 62 GRBs carried out by the ARGO-YBJ air shower detector operated in
scaler mode.
In the search for GeV gamma rays in coincidence with the
low energy GRBs detected by satellites, no evidence for VHE emission was
found for any event.
The stacked search, both in time and phase, has shown no deviation from 
the statistical expectations.
\\
Fluence upper limits as low as $\lesssim 10^{-5}$ erg cm$^{-2}$ in the 1 -- 100 GeV 
interval have been set by using ARGO-YBJ data.
Using the experimental values obtained for the GRBs with known redshift
and $\alpha$ = -2.5, 
we have calculated the corresponding minimum isotropic gamma ray energy 
$E_{iso}$.
Considering a cosmology with Hubble constant $H_0 = 70$ km~s$^{-1}$ Mpc$^{-1}$,
matter and dark energy density parameters $\Omega _m$ = 0.3 and 
$\Omega _{\Lambda}$ = 0.7, respectively \citep{Kom08}, we find the minimum 
value $E_{iso} = 2.5 \times 10^{53}$ erg for GRB071112C, emitted in the energy 
band 1.8 -- 180 GeV due to its redshift $z$=0.823.
This value is quite high, compared to the expected maximum bolometric isotropic
energy of about $10^{54}$ erg, but a beaming effect depending on the energy 
could greatly change the fraction of the total energy amount required in the 
quoted energy range.
\\
Some relevant constraints can be obtained comparing our fluence upper limits 
with the expected theoretical spectra.
Under different assumptions, the model of \citet{Asa07}
predicts GRB spectra at $z$=0.1 with fluences (in terms of $E \times F(E)$) 
roughly between $2 \times 10^{-5}$ and $2 \times 10^{-4}$ erg cm$^{-2}$ 
in our energy range of interest.
Considering the same cosmological parameters as above and the minimum
fluence upper limit of $8.1 \times 10^{-6}$ erg cm$^{-2}$ (corresponding to
GRB060801), the maximum redshift at which such a GRB could be detected
by ARGO-YBJ ranges between $z$ = 0.3 and 1.0.
\\
Since we were able to determine upper limits to the cutoff energy in the
2 -- 100 GeV energy range 
for several GRBs, we can conclude that a simple extension of the power law
spectra measured at low energies is not always possible \citep{Ban93}.
\\
Finally, the alert rate provided by the recently launched Fermi satellite, with a field of view
close to that of Swift, doubles our estimate of GRB detection \citep{Aie08}
up to a rate between 0.2 and 1 per year, and
the capability of the detector shower mode to measure the arrival direction and
energy of individual showers above a few hundred GeV allows the 
ARGO-YBJ experiment to study the GRBs in the whole 1 GeV$-$1 TeV range.




\acknowledgments

This paper is supported in part by the National Natural Science Foundation
of China (NSFC) under the grant No. 10120130794, the Chinese Ministry
of Science and Technology, the Key Laboratory of Particle Astrophysics, 
Institute of High Energy Physics (IHEP), Chinese Academy of Science (CAS), and 
the National Institute of Nuclear Physics of Italy (INFN).
\\
M. Dattoli thanks the National Institute of Astrophysics of Italy (INAF)
for partly supporting her activity in this work.

\clearpage



\clearpage

\clearpage

\clearpage

\clearpage

\clearpage

\begin{deluxetable}{lcccccrrccc}
\tabletypesize{\scriptsize}
\rotate
\tablecaption{GRBs with Measured Redshift Observed by ARGO-YBJ.
\label{tab:ul_rs}}
\tablewidth{0pt}
\tablehead{
\colhead{GRB} & \colhead{Satellite} & \colhead{$\Delta t_{90}$ (s)} &
\colhead{$\theta (^{\circ})$} & 
\colhead{Spectral Index} & \colhead{$A_{det}$ (m$^2)$} & \colhead{$\sigma$} &
\colhead{Fluence U.L. (erg cm$^{-2})$\tablenotemark{a,c}} &
\colhead{Fluence U.L. (erg cm$^{-2})$\tablenotemark{b,c}} &
\colhead{$E_{cut}$ U.L. (GeV)\tablenotemark{c}} & \colhead{Redshift $z$} \\
\colhead{(1)} & \colhead{(2)} & \colhead{(3)} &
\colhead{(4)} & \colhead{(5)} & 
\colhead{(6)} & \colhead{(7)} & \colhead{(8)} &
\colhead{(9)} &
\colhead{(10)} &
\colhead{(11)}
}
\startdata
050408 & HETE & 15 & 20.4 & CPL\tablenotemark{d} & 1560 & -2.12 & --
& 9.1$\times 10^{-5}$ & -- & 1.24 \\
050802 & Swift & 19.0 & 22.5 & 1.54 & 1516 & 0.19 & 1.0$\times 10^{-4}$
& 2.1$\times 10^{-4}$ & 8 & 1.71 \\
060115 & Swift & 139.6 & 16.6 & CPL\tablenotemark{d} & 3985 & -1.02
& -- & 7.6$\times 10^{-4}$ & -- & 3.53 \\
060526 & Swift & 298.2 & 31.7 & 2.01 & 4029 & -1.00 
& 1.8$\times 10^{-3}$ & 2.7$\times 10^{-3}$ & -- & 3.21 \\
060714 & Swift & 115.0 & 42.8 & 1.93 & 5155 & -0.61 
& 4.2$\times 10^{-3}$ & 6.9$\times 10^{-3}$ & -- & 2.71 \\
060927 & Swift & 22.5 & 31.6 & CPL\tablenotemark{d} & 5242 & -0.14 & 
-- & 5.1$\times 10^{-4}$ & -- & 5.6 \\
061110A & Swift & 40.7 & 37.3 & 1.67 & 5545 & 0.01 
& 6.7$\times 10^{-4}$ & 1.7$\times 10^{-3}$ & -- & 0.76 \\
071112C & Swift & 15 & 18.4 & 1.09 & 5198 & 1.01 
& 4.5$\times 10^{-5}$ & 1.4$\times 10^{-4}$ & $<2$ & 0.823 \\ 
081028A & Swift & 260 & 29.9 & 1.25 & 5805 & 0.37 
& 1.1$\times 10^{-3}$ & 3.0$\times 10^{-3}$ & 4 & 3.038 \\ 
\enddata
\tablecomments{$^a$Using the spectrum determined by satellites.
$^b$Assuming a differential spectral index 2.5.
$^c$99\% c.l..
$^d$See text.}
\end{deluxetable}

\begin{deluxetable}{lccccrrccc}
\tabletypesize{\scriptsize}
\rotate
\tablecaption{GRBs with no Redshift ($z$ = 1 is assumed) Observed by 
ARGO-YBJ.
\label{tab:ul}}
\tablewidth{0pt}
\tablehead{
\colhead{GRB} & \colhead{Satellite} & \colhead{$\Delta t_{90}$ (s)} &
\colhead{$\theta (^{\circ})$} & \colhead{Spectral Index} & 
\colhead{$A_{det}$ (m$^2)$} & \colhead{$\sigma$} &
\colhead{Fluence U.L. (erg cm$^{-2})$\tablenotemark{a,c}} &
\colhead{Fluence U.L. (erg cm$^{-2})$\tablenotemark{b,c}} &
\colhead{$E_{cut}$ U.L. (GeV)\tablenotemark{c}} \\
\colhead{(1)} & \colhead{(2)} & \colhead{(3)} &
\colhead{(4)} & \colhead{(5)} & 
\colhead{(6)} & \colhead{(7)} & \colhead{(8)} &
\colhead{(9)} & \colhead{(10)} 
}
\startdata
041228 & Swift & 55.4 & 28.1 & 1.60 & 563 & -0.01 & 4.4$\times 10^{-4}$ 
& 1.1$\times 10^{-3}$ & 38 \\
050509A & Swift & 11.4 & 34.0 & 2.11 & 1473 & 0.62 & 1.9$\times 10^{-4}$ 
& 3.0$\times 10^{-4}$ & -- \\
050528 & Swift & 11.3 & 37.8 & 2.27 & 1473 & 0.71 & 8.0$\times 10^{-4}$ 
& 1.0$\times 10^{-3}$ & -- \\
051105A & Swift & 0.1 & 28.5 & 1.22 & 3119 & 1.24 & 1.6$\times 10^{-5}$ 
& 4.8$\times 10^{-5}$ & 6 \\
051114 & Swift & 2.2 & 32.8 & 1.21 & 3032 & 3.37 & 5.6$\times 10^{-5}$ 
& 1.8$\times 10^{-4}$ & 5 \\
051227 & Swift & 114.6 & 22.8 & 1.45 & 2989 & 0.44 & 3.7$\times 10^{-4}$ 
& 9.2$\times 10^{-4}$ & 27 \\
060105 & Swift & 54.4 & 16.3 & 1.07 & 3119 & 1.77 & 2.5$\times 10^{-4}$ 
& 7.7$\times 10^{-4}$ & $<2$ \\
060111A & Swift & 13.2 & 10.8 & CPL\tablenotemark{d} & 3206 & 0.39 & -- & 
9.7$\times 10^{-5}$ & -- \\
060121 & HETE & 2.0 & 41.9 & Band\tablenotemark{d} & 4159 & 0.60 & 
2.5$\times 10^{-4}$ & 2.8$\times 10^{-4}$ & -- \\
060421 & Swift & 12.2 & 39.3 & 1.55 & 3855 & -0.51 & 2.0$\times 10^{-4}$ 
& 5.2$\times 10^{-4}$ & 29 \\
060424 & Swift & 37.5 & 6.7 & 1.71 & 4072 & 0.12 & 7.5$\times 10^{-5}$ 
& 1.6$\times 10^{-4}$ & -- \\
060427 & Swift & 64.0 & 32.6 & 1.87 & 4115 & -0.13 & 2.6$\times 10^{-4}$ 
& 5.1$\times 10^{-4}$ & -- \\
060510A & Swift & 20.4 & 37.4 & 1.57 & 3899 & 2.42 & 6.8$\times 10^{-4}$ 
& 1.8$\times 10^{-3}$ & 21 \\
060717 & Swift & 3.0 & 7.4 & 1.70 & 5155 & 1.58 & 2.3$\times 10^{-5}$ 
& 4.8$\times 10^{-5}$ & -- \\
060801 & Swift & 0.5 & 16.8 & 0.47 & 5415 & 0.81 & 8.1$\times 10^{-6}$ 
& 3.0$\times 10^{-5}$ & $<2$ \\
060805B & IPN & 8 & 29.1 & Band\tablenotemark{d} & 5285 & -0.45 & 
1.2$\times 10^{-4}$ & 1.1$\times 10^{-4}$ & -- \\
060807 & Swift & 54.0 & 12.4 & 1.58 & 5155 & 0.78 & 1.3$\times 10^{-4}$ 
& 3.0$\times 10^{-4}$ & 34 \\
061028 & Swift & 106.2 & 42.5 & 1.73 & 5458 & -3.33 & 3.5$\times 10^{-4}$ 
& 8.0$\times 10^{-4}$ & -- \\
061122 & INTEGRAL & 18 & 33.5 & CPL\tablenotemark{d} & 5025 & 0.60 & -- & 
6.4$\times 10^{-4}$ & -- \\
070201 & IPN & 0.3 & 20.6 & CPL\tablenotemark{d} & 5242 & -1.21 & -- & 
1.2$\times 10^{-5}$ & -- \\
070219 & Swift & 16.6 & 39.3 & 1.78 & 4982 & -0.71 & 3.1$\times 10^{-4}$ 
& 6.8$\times 10^{-4}$ & -- \\
070306 & Swift & 209.5 & 19.9 & 1.66 & 2513 & -0.83 & 5.4$\times 10^{-4}$ 
& 1.2$\times 10^{-3}$ & 73 \\
070531 & Swift & 44.5 & 44.3 & 1.41 & 2816 & 0.59 & 6.6$\times 10^{-4}$ 
& 1.9$\times 10^{-3}$ & 24 \\
070615 & IINTEGRAL & 30 & 37.6 & -- & 5328 & 1.81 & -- & 1.7$\times 10^{-3}$ 
& -- \\
071013 & Swift & 26 & 13.3 & 1.60 & 4765 & -0.06 & 5.2$\times 10^{-5}$ 
& 1.2$\times 10^{-4}$ & 48 \\
071101 & Swift & 9.0 & 32.8 & 2.25 & 3596 & 1.01 & 1.6$\times 10^{-4}$ 
& 2.1$\times 10^{-4}$ & -- \\
071104 & AGILE & 12 & 19.9 & -- & 4029 & -0.07 & -- & 1.3$\times 10^{-4}$ & -- \\
071118 & Swift & 71 & 41.2 & 1.63 & 5025 & 0.54 & 1.3$\times 10^{-3}$ 
& 3.3$\times 10^{-3}$ & -- \\
080328 & Swift & 90.6 & 37.2 & 1.52 & 6065 & -1.19 & 7.6$\times 10^{-4}$ 
& 2.1$\times 10^{-3}$ & 13 \\
080602 & Swift & 74 & 42.0 & 1.43 & 5762 & 1.24 & 1.1$\times 10^{-3}$ 
& 3.1$\times 10^{-3}$ & 18 \\
080613B & Swift & 105 & 39.2 & 1.39 & 5718 & 0.65 & 1.2$\times 10^{-3}$ 
& 3.6$\times 10^{-3}$ & 10 \\
080727C & Swift & 79.7 & 34.5 & CPL\tablenotemark{d} & 5415 & 3.52 & -- & 
1.4$\times 10^{-3}$ & -- \\
080822B & Swift & 64 & 40.3 & 2.54 & 5762 & -1.84 & 1.3$\times 10^{-3}$ &
1.3$\times 10^{-3}$ & -- \\
080830 & Fermi & 45 & 37.1 & Band\tablenotemark{d} & 5805 & -0.04 & 
6.3$\times 10^{-4}$ & 1.5$\times 10^{-3}$ & -- \\
080903 & Swift & 66 & 21.5 & CPL\tablenotemark{d} & 5588 & -1.33 & -- & 
2.3$\times 10^{-4}$ & -- \\
081025 & Swift & 23 & 30.5 & 1.12 & 5718 & -0.48 & 6.0$\times 10^{-5}$ & 
2.0$\times 10^{-4}$ & $<2$ \\
081102B & Fermi & 2.2 & 27.8 & 1.07 & 5762 & 0.02 & 1.7$\times 10^{-5}$ & 
5.8$\times 10^{-5}$ & $<2$ \\
081105 & IPN & 10 & 36.7 & -- & 5718 & -0.77 & -- & 
4.0$\times 10^{-4}$ & -- \\
081122 & Fermi & 26 & 8.3 & Band\tablenotemark{d} & 4289 & -2.03 & 
5.5$\times 10^{-5}$ & 7.1$\times 10^{-5}$ & -- \\
081128 & Swift & 100 & 31.8 & CPL\tablenotemark{d} & 5242 & -0.63 & -- & 
9.8$\times 10^{-4}$ & -- \\
081130B & Fermi & 12 & 28.6 & CPL\tablenotemark{d} & 5978 & -0.05 & -- & 
2.2$\times 10^{-4}$ & -- \\
081215A & Fermi & 7.7 & 35.9 & Band\tablenotemark{d} & 5762 & -0.15 & 
3.1$\times 10^{-4}$ & 5.1$\times 10^{-4}$ & -- \\
090107A & Swift & 12.2 & 40.1 & 1.69 & 5762 & -1.12 & 2.2$\times 10^{-4}$ & 
5.3$\times 10^{-4}$ & -- \\
090118 & Swift & 16 & 13.4 & 1.35 & 5805 & -1.62 & 2.1$\times$10$^{-5}$ & 
5.5$\times$10$^{-5}$ & 3 \\
090126B & Fermi & 10.8 & 3.7 & CPL\tablenotemark{d} & 5892 & -1.43 & -- &
4.0$\times$10$^{-5}$ & -- \\ 
090227B & Fermi & 0.9 & 9.7 & CPL\tablenotemark{d} & 5935 & 0.21 & -- &
1.6$\times$10$^{-5}$ & -- \\
090301 & Swift & 41.0 & 14.2 & CPL\tablenotemark{d} & 5805 & 0.73 & -- &
2.3$\times$10$^{-4}$ & -- \\
090301B & Fermi & 28 & 24.3 & Band\tablenotemark{d} & 5892 & -2.20 & 
6.2$\times$10$^{-5}$ & 1.1$\times$10$^{-4}$ & -- \\
090306B & Swift & 20.4 & 38.5 & CPL\tablenotemark{d} & 5805 & -0.65 & -- &
9.3$\times$10$^{-4}$ & -- \\
090320B & Fermi & 52 & 29.0 & CPL\tablenotemark{d} & 5892 & -0.25 & -- &
2.4$\times$10$^{-4}$ & -- \\
090328B & Fermi & 0.32 & 15.5 & Band\tablenotemark{d} & 5848 & 0.48 & 
1.7$\times$10$^{-5}$ & 1.8$\times$10$^{-5}$ & -- \\
090403 & Fermi & 16 & 28.5 & -- & 6021 & 0.65 & -- &
2.5$\times$10$^{-4}$ & -- \\
090407 & Swift & 310 & 45.0 & 1.73 & 6021 & 1.53 & 5.0$\times$10$^{-3}$ & 
1.1$\times$10$^{-2}$ & -- \\

\enddata
\tablecomments{$^a$Using the spectrum determined by satellites.
$^b$Assuming a differential spectral index 2.5.
$^c$99\% c.l..
$^d$See text.}
\end{deluxetable}


\end{document}